\documentclass[10pt, conference, compsocconf]{IEEEtran}
\pdfoutput=1

\usepackage{pdfpages}
\usepackage{hyperref}
\usepackage[absolute]{textpos}
\usepackage{url}
\usepackage{soul} % Package for word/sentence highlighting
\usepackage{amsmath}
\usepackage{amssymb} % Package with an extended symbol collection.
\usepackage{cite} % Package for citations.
\usepackage{graphicx} % Package for graphics.
\usepackage{multirow}
\usepackage{marginnote} % for margin notes
\usepackage{wrapfig}

\DeclareRobustCommand{\hlcyan}[1]{{\sethlcolor{cyan}\hl{#1}}}

\usepackage[linesnumbered, boxruled]{algorithm2e}
\SetKwInput{Algorithm}{Algorithm}
\SetKwInput{Result}{Result}

\graphicspath{figures/}
\renewcommand{\baselinestretch}{0.95}

\newcommand{\stitle}[1]{\vspace{1ex}\noindent\textbf{#1}}

\IEEEoverridecommandlockouts

\begin{document}
\begin{textblock}{24}(1,0.2)
    \noindent\tiny  This paper is a preprint; it has been published in 2019 IEEE World Congress on Services (SERVICES), Workshop on Cyber Security \& Resilience in the Internet of Things (CSRIoT @ IEEE Services), Milan, Italy, July 2019   \\
    \textbf{IEEE copyright notice} \textcopyright 2021 IEEE. Personal use of this material is permitted. Permission from IEEE must be obtained for all other uses, in any current or future media, including reprinting/republishing this material for advertising or promotional purposes,\\ creating new collective works, for resale or redistribution to servers or lists, or reuse of any copyrighted component of this work in other works.
    \end{textblock}

\title{A crawler architecture for harvesting the clear, social, and dark web\\ for IoT-related cyber-threat intelligence%$^*$
\thanks{
\protect\begin{wrapfigure}[3]{l}{0.9cm}
\protect\raisebox{-12.5pt}[0pt][7pt]{\protect\includegraphics[height=.9cm]{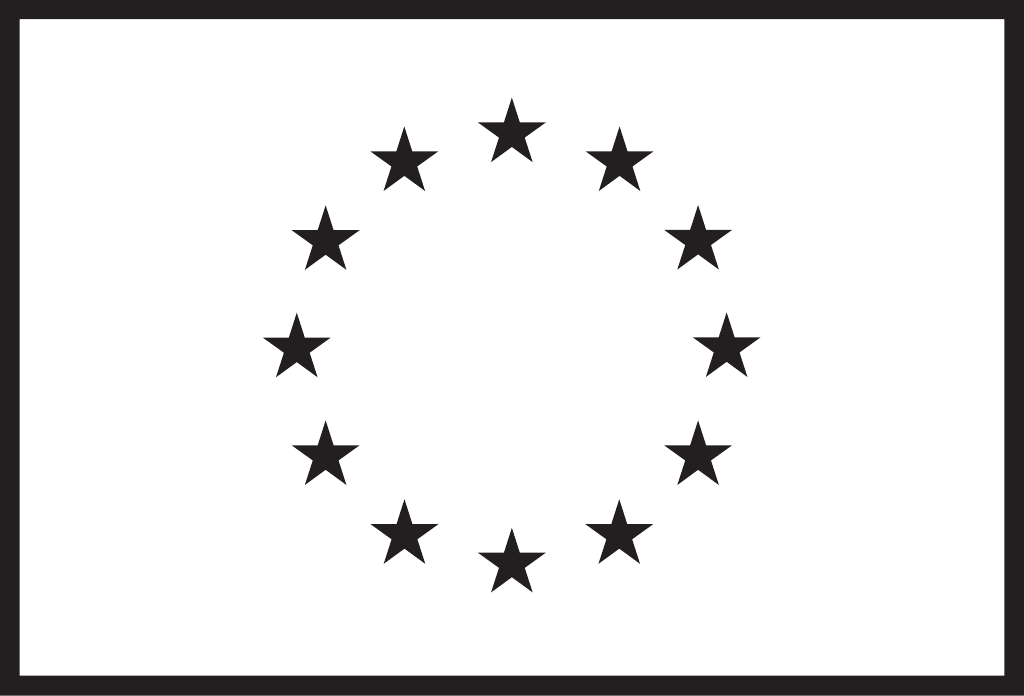}}
\protect\end{wrapfigure}
	This work has received funding from the European Union's Horizon 2020 research and innovation programme under grant agreement no. 786698. The work reflects only the authors' view and the Agency is not responsible for any use that may be made of the information it contains.
	}
} 

\author{\IEEEauthorblockN{Paris Koloveas\ \ \ \ \  Thanasis Chantzios\ \ \ \ \  Christos Tryfonopoulos\ \ \ \ \  Spiros Skiadopoulos}
\IEEEauthorblockA{University of the Peloponnese, GR22131, Tripolis, Greece\\
\{pkoloveas, tchantzios, trifon, spiros\}@uop.gr}
}

%\vspace{-20mm}
\maketitle

% !TEX root = main.tex

\begin{abstract}
The clear, social, and dark web have lately been identified as rich sources of valuable cyber-security information that --given the appropriate tools and methods-- may be identified, crawled and subsequently leveraged to actionable cyber-threat intelligence. In this work, we focus on the information gathering task, and present a novel crawling architecture for transparently harvesting data from security websites in the clear web, security forums in the social web, and hacker forums/marketplaces in the dark web. The proposed architecture adopts a two-phase approach to data harvesting. Initially a machine learning-based crawler is used to direct the harvesting towards websites of interest, while in the second phase state-of-the-art statistical language modelling techniques are used to represent the harvested information in a latent low-dimensional feature space and rank it based on its potential relevance to the task at hand. % to contain exploitable cyber-security information.
The proposed architecture is realised using exclusively open-source tools, and a preliminary evaluation with crowdsourced results demonstrates its effectiveness.
\end{abstract}


% \IEEEpubid{0000--0000/00\$00.00~\copyright~200X IEEE}

% !TEX root = main.tex

\section{Introduction}
\label{sec:intro}

Over the years cyber-threats have increased in numbers and sophistication; adversaries now use a vast set of tools and tactics to attack their victims with their motivations ranging from intelligence collection to destruction or financial gain. 
%This sentence could go away...
%Organizations worldwide, from governments to public and corporate enterprises, are under constant threat by these evolving cyber-attacks. 
%
Lately, the utilisation of IoT devices on a number of applications, ranging from home automation to monitoring of critical infrastructures, has created an even more complicated cyber-defense landscape. The sheer number of IoT devices deployed globally, most of which are readily accessible and easily hacked, allows threat actors to use them as the cyber-weapon delivery system of choice in many today’s cyber-attacks, ranging from botnet-building for DDoS attacks, to malware spreading and spamming. 

Trying to stay on top of these evolving cyber-threats has become an increasingly difficult task, and timeliness in the delivery of relevant cyber-threat related information is essential for appropriate protection and mitigation. Such information is typically leveraged from collected data, and includes zero-day vulnerabilities and exploits, indicators (system artefacts or observables associated with an attack), security alerts, threat intelligence reports, as well as recommended security tool configurations, and is often referred to as \textit{cyber-threat intelligence} (CTI). To this end, with the term CTI we typically refer to any information that may help an organization identify, assess, monitor, and respond to cyber-threats. In the era of big data, it is important to note that the term intelligence does not typically refer to the data itself, but rather to information that has been \textit{collected, analysed, leveraged} and \textit{converted} to a series of actions that may be followed upon, i.e., has become \textit{actionable}.

While CTI may be collected by resorting to a variety of means (e.g., monitoring cyber-feeds) and from a variety of sources, we are particularly interested in gathering CTI from the clear, social, and dark web where threat actors collaborate, communicate and plan cyber-attacks. Such an approach allows us to provide visibility to a number of sources that are of preference to threat-actors and identify timely CTI including zero-day vulnerabilities and exploits. To do so, we envision an integrated framework that encompasses key technologies for pre-reconnaissance CTI \textit{gathering, analysis} and \textit{sharing} through the use of state-of-the-art tools and technologies. In this context, newly discovered data from various sources will be inspected for their relevance to the task (\textit{gathering}), and discovered CTI in the form of vulnerabilities, exploits, threat actors, or cyber-crime tools will be identified (\textit{analysis}) and stored in a vulnerability database using existing formats like CVE\footnote{\url{https://www.mitre.org}} and CPE\footnotemark[1] (\textit{sharing}).

\begin{figure*}[t]
	\centering
	\includegraphics[scale=0.55]{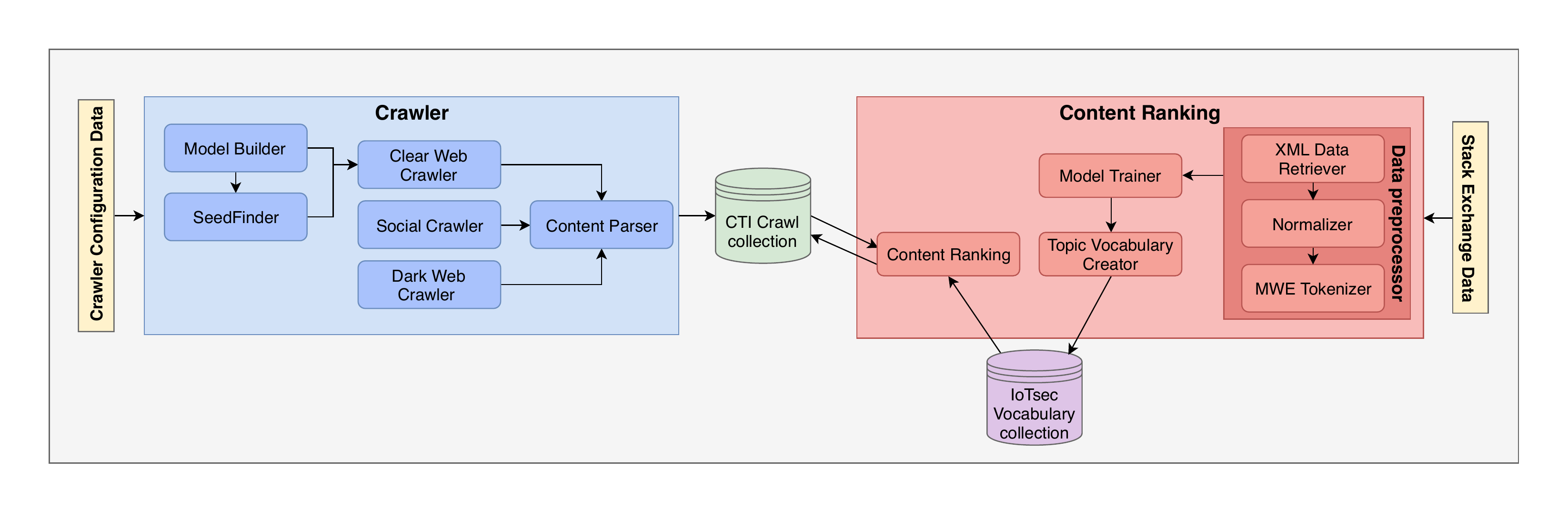}
	\vspace{-12mm}
	\caption{A high-level view of the proposed architecture}
	\label{fig:completeArchitecture}
	\vspace{-6mm}
\end{figure*}

In this work we focus on the \textit{gathering} part of the envisioned framework, and present a novel architecture that is able to transparently provide a crawling infrastructure for a variety of CTI sources in the clear, social, and dark web. Our approach employs a thematically focused crawler for directing the crawl towards websites of interest to the CTI gathering task. This is realised by resorting to a combination of machine learing techniques (for open domain crawl) and regex-based link filtering (for structured domains like forums). The retrieved content is stored in an efficient NoSQL datastore and is retrieved for further inspection in order to decide its usefulness to the task. This is achieved by employing statistical language modelling techniques \cite{MikolovW2V} to represent all information in a latent low-dimensional feature space and a ranking-based approach to the collected content (i.e., rank it according to its potential to be useful). These techniques allow us to train our language model to  (i) capture and exploit the most salient words for the given task by building upon user conversations, (ii) compute the semantic relatedness between the crawled content and the task at hand by leveraging the identified salient words, and (iii) classify the crawled content according to its relevance/usefulness based on its semantic similarity to CTI gathering. Notice that the post-mortem inspection of the crawled content is necessary, since the thematically focused crawl is forced to make a crude decision on the link relevance (and if it should be visited or not) since it resorts on a limited feature space (e.g., alt-text of the link, words in the url, or relevance of the parent page).

To the best of our knowledge, this is the first approach to CTI gathering that views the crawling task as a two-stage process, where a crude classification is initially used to prune the crawl frontier, while a more refined approach based on the collected content is used to decide on its relevance to the task. This novel approach to CTI gathering is integrated to an infrastructure that is entirely open-source and able to transparently monitor the clear, social, and dark web. 

The rest of the paper is organised as follows. In the next section, we present the architecture and provide details on several design and implementation choices, while in Section~\ref{sec:eval} we present our evaluation plan and a preliminary effectiveness evaluation using both crowdsourced results and anecdotal examples. Finally, Section~\ref{sec:relatedWork} outlines related work, while Section~\ref{sec:outlook} discusses future research directions.

%meetCyber-Threat Intelligence  is gathered by crawling and searching a variety of sources (Surface, Deep and Dark Web). After the relevant web pages are collected, they are temporary stored for further processing. In the general case, the collected pages can be viewed as text files. This textual information should be further processed in order to extract the relevant CTI information. This is a complex process that involves natural language processing, text classification, entity extraction and resolution and ontology integration and mapping. Given the evolving nature of current cyber-threats and the inherent distribution of the relevant information sources, the task of collecting, identifying, mining, leveraging, and sharing actionable cyber-threat intelligence has become increasingly complex. 

%Therefore, the main challenge organizations face is the abundance of data and the lack of actionable intelligence.

% !TEX root = main.tex
\section{Architecture}
\label{sec:architecture}

The proposed architecture consists of two major components: the \emph{crawling module} and the \emph{content ranking module}. The idea behind this two-stage approach is the openness of the topic at hand that cannot be accurately modelled by a topical crawler, which mainly focuses on staying within the given topic. The difficulty emerges from websites that, although relevant to the topic (e.g., discussing IoT security in general), have no actual information that may be leveraged to actionable intelligence (e.g., do not mention any specific IoT related vulnerability). To overcome this challenge, a focused crawler that employs machine learning techniques to direct the crawl is aided by advanced language models to decide on the usefulness of the crawled websites by utilising the harvested content. To this end, we designed a novel module that employs a ranking-based approach to the assessment of the usefulness of a website content using language models. To capture the background knowledge regarding vocabulary and term correlations, we resorted to latent topic models \cite{MikolovW2V}, while ranking is modelled as vector similarity to the task. 

The proposed architecture has been entirely designed on and developed using open-source software; it employs an open-source focused crawler\footnote{\url{https://github.com/ViDA-NYU/ache}}, an open source implementation of word embeddings\footnote{\url{https://radimrehurek.com/gensim/}} for the latent topic modeling, and an open-source NoSQL database\footnote{\url{https://www.mongodb.com/}} for the storage of the topic models and the crawled content. The front-end modules are based on  HTML, CSS and JavaScript. Figure \ref{fig:completeArchitecture} displays the complete architecture with all the sub-components that will be detailed in the following sections. 

\subsection{Clear/social/dark web crawling}
\label{sec:crawler}

The crawling module is based upon the open-source implementation of the ACHE Crawler\footnotemark[2] and contains three different sub-components: (i) a focused crawler for the clear web, (ii) an in-depth crawler for the social web (i.e., forums), and (iii) a TOR-based crawler for the dark web. Below, we outline the characteristics of the different sub-components and present the rationale behind our implementation choices.

%\subsubsection{\textbf{Clear Web Crawler}}
%\label{sec:clear-web-crawler}

\stitle{Clear web crawler.} This sub-component is designed to perform \emph{focused crawls} on the clear web with the purpose to discover new resources that may contain cyber-threat intelligence. To direct the crawl towards topically relevant websites we utilise an SVM classifier, which is trained by resorting on an equal number of positive and negative examples of websites that are used as input to the \emph{Model Builder} component of ACHE. Table~\ref{tab:pos_neg_examples} contains some example entries that are specific to our task. Subsequently, the \emph{SeedFinder}\cite{VieiraSeedFinder} component is utilised to aid the process of locating initial seeds for the focused crawl on the clear web; this is achieved by combining the classification model built previously with a user-provided query relevant to the topic - in our case we use the query ``iot vulnerabilities''. 

\begin{table} [t]
	\caption{Positive \& Negative Webpage Examples}
	\vspace{-3mm}
	\centering
	\begin{tabular}{|p{4cm}|p{4cm}|}
		\hline
		\textbf{Positive} 	&  \textbf{Negative}	\\	\hline\hline
		IoT, Cloud, or Mobile: All Ripe for Exploit and Need Security’s Attention $\vert$ CSO Online	&	Scammers pose as CNN's Wolf Blitzer, target security professionals $\vert$ CSO Online	
%		\\ \hline
%		\url{https://www.csoonline.com/article/3119765/security/hackers-found-47-new-vulnerabilities-in-23-iot-devices-at-def-con.html}			&	\url{https://www.csoonline.com/article/3304306/network-security/what-is-an-rdp-attack-7-tips-for-mitigating-your-exposure.html}	
		\\ \hline
		New IoT Threat Exploits Lack of Encryption in Wireless Keyboards $\vert$ eSecurity Planet		&
		19 top UEBA vendors to protect against insider threats and external attacks $\vert$ eSecurity planet	
		\\ \hline
		Security Testing the Internet of Things: Dynamic testing (Fuzzing) for IoT security $\vert$ Beyond Security	& 
		Build your own cloud $\vert$ SoftLayer
			
%		\\ \hline
%		\url{https://securingtomorrow.mcafee.com/mcafee-labs/insight-into-home-automation-reveals-vulnerability-in-simple-iot-product}			&	\url{https://securingtomorrow.mcafee.com/mcafee-labs/cactustorch-fileless-threat-abuses-net-to-infect-victims/}	
%		\\ \hline
%		\url{https://www.bleepingcomputer.com/news/security/passwords-for-tens-of-thousands-of-dahua-devices-cached-in-iot-search-engine/}		&	\url{https://www.bleepingcomputer.com/news/security/malware-disguised-as-job-offers-distributed-on-freelance-sites/}	
%		\\ \hline
%		\url{https://www.bleepingcomputer.com/news/security/half-a-billion-iot-devices-vulnerable-to-dns-rebinding-attacks/}					&	\url{https://www.bleepingcomputer.com/news/security/kraken-cryptor-ransomware-masquerading-as-superantispyware-security-program/}	
		\\ \hline
	\end{tabular}
	\label{tab:pos_neg_examples}
	\vspace{-7mm}
\end{table}

%\subsubsection{\textbf{Forum Crawler}}
%\label{sec:forum-crawler}

\stitle{Social web crawler.} This sub-component is used to perform \emph{in-depth crawls} on specific selected forums on the social web,  where the topic of discussion matches the given task. To this end, the social web crawler can be provided with links to discussion threads on IoT vulnerabilities and use them to traverse the forum structure and download all relevant discussions on the topic. Notice that contrary to clear web crawling, in the forum crawl all links are considered relevant by default since they correspond to discussions over the given topic, so there is no need for utilising a page classification model. However, to filter out parts of the forum that are irrelevant or non-informative (e.g., user info pages), we employ \emph{regex-based link filters} that may be applied within specific forums or in a cross-forum fashion. Table~\ref{tab:link_filters} contains some example link filters. Notice that this type of crawl is primarily used for thread/forum monitoring and is not targeted in identifying new websites.

%\subsubsection{\textbf{Deep/Dark Web Crawler}}
%\label{sec:deep-dark-crawler}

\stitle{Dark web crawler.} This sub-component is used to perform \emph{in-depth crawls} on specific websites on the dark web by utilising TOR proxies. To do so, the crawler is provided with a number of onion links that correspond to hacker forums or marketplaces selling cybercrime tools and zero-day vulnerabilities/exploits and monitors the discussions for content of interest. To overcome user authentication mechanisms that are often in place in dark web forums/marketplaces, the dark web crawler requires an initial manual login. After a successful user authentication, the session cookies are stored and are utilised (via HTTP requests) in subsequent visits of the crawler to simulate user login. After each crawl completes a set interval, the crawled HTML pages are parsed by the \emph{content parser} sub-component, which extracts the textual content along with useful metadata (e.g., bitcoin value of sold cybercrime tools or user fame/activity/reputation level).

All content from the different (clear/social/dark) web crawling components is downloaded in its raw HTML format and stored in a NoSQL document store (mongoDB\footnotemark[4]) for further processing as discussed in the next section.

\subsection{Content ranking and classification}
\label{sec:classifier}

Deciding whether a crawled website contains useful cyber-threat intelligence is a challenging task given the typically generic nature of many websites that discuss general security issues. To tackle this problem, we designed and implemented a novel content ranking module that assesses the relevance and usefulness of the crawled content. To do so, we represent the topic as a vocabulary distribution by utilising distributional vectors of related words; for example a topic on IoT security could be captured by related words and phrases like ``Mirai botnet'', ``IoT'', or ``exploit kits''. Such salient phrases related to the topic may be obtained by un-/semi-supervised training of latent topic models over external datasets such as IoT and security related forums. In this way, we are able to capture semantic dependencies and statistical correlations among words for a given topic and represent them in a low-dimension latent space by state-of-the-art latent topic models \cite{MikolovW2V}. 

Since useful cyber-threat intelligence manifests itself in the form of cyber-security articles, user posts in security/hacker forums, or advertisement posts in cybercrime marketplaces, it can also be characterised as distributional vectors of salient words. Then, the similarity between the distributional vectors of harvested content and the given topic (i.e., IoT vulnerabilities) may be used to assess the content relevance to the topic.

%For the Classifier we used Word2Vec; a statistical model for \emph{Efficient Estimation of Word Representations in Vector Space} proposed by Mikolov et al. \cite{MikolovW2V} The idea behind using Word2Vec, was that we could find the co-occurring words in a dataset, and by looking at specific words that are relevant to our topic, we could create a \emph{Topic Vocabulary}. The first problem we had to solve was to create an appropriate corpus for this task. 

\begin{table} [t]
	\caption{Regex-based link filters}
	\vspace{-3mm}
	\centering
	\begin{tabular}{|l|p{3.45cm}|p{2.8cm}|}
		\hline
		Category & Pattern & Operation \\
		\hline\hline
		Whitelist & \url{https://www.wilderssecurity.com/threads/*} & Crawl content only from the \emph{\textbf{Threads}} section of the domain	\\	\hline
		Whitelist & \url{https://blogs.oracle.com/security/*} & Crawl all articles about \emph{\textbf{Security}}						\\	\hline
		Blacklist & \url{https://www.wilderssecurity.com/members/*} 	& Crawl the entire domain apart from the \emph{\textbf{Members Area}}		\\	\hline
		Blacklist & \url{https://www.securityforum.org/events/*} 	& Crawl the entire domain apart from \emph{\textbf{Events}} 				\\	\hline
	\end{tabular}
	\vspace{-7mm}
	\label{tab:link_filters}
\end{table}

To better capture the salient vocabulary that is utilised by users for the IoT security domain we resorted to a number of different discussion forums within the Stack Exchange ecosystem to create a training dataset. To this end, we utilised the \emph{Stack Exchange Data Dump}\footnote{\url{https://archive.org/details/stackexchange}} to get access to IoT and information security related discussion forums including \textit{Internet of Things}, \textit{Information Security}, \textit{Arduino}, \textit{Raspberry Pi}, and others. The utilised data dumps contain user discussions in Q\&A form, including the text from \emph{posts}, \emph{comments} and related \emph{tags}, and were used as input to the \textit{data preprocessor} sub-component described below.
%\footnote{https://iot.stackexchange.com/}
%\footnote{https://security.stackexchange.com/}
%\footnote{https://arduino.stackexchange.com/}
%\footnote{https://raspberrypi.stackexchange.com/}

%The last two were selected because they are the most prominent devices for custom IoT projects with very active communities, so their data would help our model to understand the IoTSec vocabulary even more. \\

%In the following sections we are going to outline the 4 major components that are part of the \emph{Classifier} component.

%\subsubsection{\textbf{Preprocessor}} 
%\label{sec:corpus-creator}

\stitle{Data preprocessor.} This sub-component is responsible for the data normalisation process that involves a number of steps described in the following. The input data for the sub-component is typically XML-formatted and at first an XML DOM parser is used to parse the data and keep only the useful part that contains user posts, comments, and tags. The parsed data are, subsequently, fed into the \emph{Normalizer} that performs typical normalisation (e.g., case folding, symbol removal) and anonymisation (e.g., username elimination) actions. Finally, the third step in the preprocessing phase (\emph{Multi-Word Expression tokenisation}) includes the identification and characterisation of important multi-word terms (such as ``exploit kits'' or ``Mirai botnet'') in order to extend the functionality of the skip-gram model \cite{MikolovW2V} for such terms. 

% we use the gensim implementation, not the one on the footnote - 
\stitle{Model trainer.} The preprocessed document corpus is subsequently utilised to train the language model \cite{MikolovW2V}; this is done by using the Gensim\footnotemark[3] open source implementation of word2vec (based on a neural network for word relatedness), setup with a latent space of 150 dimensions, a training window of 5 words, a minimum occurrence of 1 term instance, and 10 parallel threads. The result of the training is a 150-dimensions distributional vector for each term that occurs at least once in the training corpus.

% DDoS user tags?
\stitle{Topic vocabulary creator.} To automatically extract the set of salient words that will be used to represent the topic we utilised the extracted user tags, and augmented them with the set of $N$ most related terms in the latent space for each user tag; term relatedness was provided by the trained language models and the corresponding word vectors. Table \ref{tab:closest_topic_words} shows an example of the most relevant terms to the DDoS user tag, for $N=$5,10, and 15. The resulting (expanded) vocabulary is stored in a separate NoSQL document store (mongoDB\footnotemark[4]).

%\begin{table} [t]
%	\caption{Most Relevant Terms for Tag: \textbf{DDoS}}
%	\centering
%	\begin{tabular}{|c|c|}
%		\hline
%		\textbf{word} 	&  \textbf{score}		\\	\hline\hline
%		dos 			& 0.8098790645599365	\\	\hline
%		volumetric		& 0.7910763025283813	\\	\hline
%		flooding		& 0.7676426172256470	\\	\hline
%		cloudflare		& 0.7591973543167114	\\	\hline
%		prolexic 		& 0.7564446926116943	\\	\hline
%		flood 			& 0.7523617744445801	\\	\hline
%		aldos 			& 0.7459267377853394	\\	\hline
%		floods 			& 0.7398521304130554	\\	\hline
%		ip\_spoofing 	& 0.7388113141059875	\\	\hline
%		radware 		& 0.7373002767562866	\\	\hline
%		slowloris		& 0.7246818542480469	\\	\hline	
%		botnet 			& 0.7212058305740356	\\	\hline
%		drdos 			& 0.7181754112243652	\\	\hline
%		blackholing		& 0.7169653177261353	\\	\hline
%		amplification 	& 0.7142491340637207	\\	\hline
%	\end{tabular}
%	\label{tab:closest_topic_words}
%\end{table}

\begin{table} [t]
	\caption{Most relevant terms for tag DDoS}
	\vspace{-3mm}
	\centering
	\begin{tabular}{|c|c||c|c||c|c|}
		\hline
		rank &  term  & rank & term & rank & term\\		\hline\hline
		\#1  & dos & \#6 & flood & \#11 & slowloris\\
		\hline
		\#2 & volumetric	 & \#7 & aldos & \#12 & botnet\\
		\hline
		\#3 & flooding & \#8 & floods & \#13 & drdos\\
		\hline
		\#4 & cloudflare & \#9 & ip spoofing & \#14 & blackholing\\
		\hline
		\#5 & prolexic & \#10 & radware & \#15 & amplification\\
		\hline
	\end{tabular}
	\label{tab:closest_topic_words}
	\vspace{-7mm}
\end{table}

\stitle{Content ranking.} To assess the relevance and usefulness of the crawled content we employ the \emph{content ranking} sub-component; this component utilises the expanded vocabulary created in the previous phase to decide how similar a crawled post is to the topic by computing the similarity between the topic and post vectors. This is done as follows.

The topic vector $\vec{T}$ is constructed as the sum of the distributional vectors of all the topic terms $\vec{t_i}$ that exist in the topic vocabulary, i.e.,  
\[\vec{T} = \sum\nolimits_{\forall i}{\vec{t_i}}\]
Similarly, the post vector $\vec{P}$  is constructed as the sum of the distributional vectors of all the post terms $\vec{w_j}$ that are present in the topic vocabulary. To promote the impact of words related to the topic at hand, we introduce a topic-dependent weighting scheme for post vectors in the spirit of \cite{sigir16-BGMMTW}. Namely for a topic $T$ and a post containing the set of words $\{w_1, w_2, \ldots \}$, the post vector is computed as
\[\vec{P} = \sum\nolimits_{\forall j}{\cos{(\vec{w_j}, \vec{T}})}\cdot\vec{w_j}\]

Finally, after both vectors have been computed, the relevance score $r$ between the topic $T$ and a post $P$ is computed as the cosine similarity of their respective distributional vectors in the latent space
\[r = \cos{(\vec{T}, \vec{P})}\]

Having computed a relevance score for every crawled post in the NoSQL datastore, the classification task of identifying relevant/useful posts is trivially reduced to either a thresholding or a top-k selection operation. Notice that the most/least relevant websites may also be used to reinforce the crawler model (i.e., as input to the \textit{Model Builder} sub-component). Comparing the pros and cons of thresholding and top-k post selection in the context of assessing a continuous crawling task is an ongoing research effort. 
% !TEX root = main.tex

\section{Preliminary Evaluation}
\label{sec:eval}

\begin{figure*}[t]
	\centering
	\includegraphics[scale=0.55]{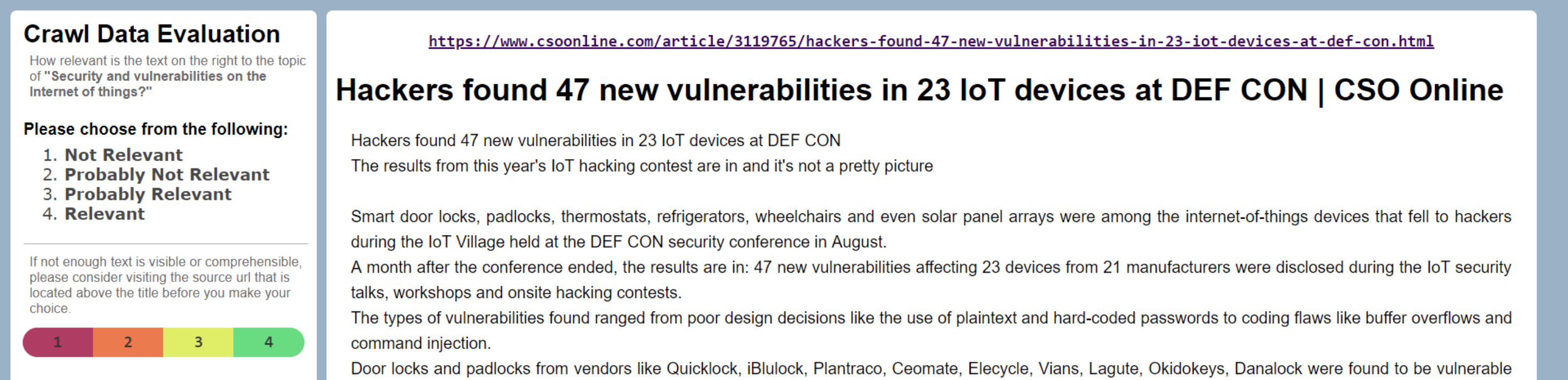}
	\vspace{-2mm}
	\caption{Web-based evaluation tool for collecting crowdsourced data from human experts}
	\label{fig:CrawledDataEvaluation}
	\vspace{-4mm}
\end{figure*}

In this section we present a preliminary evaluation of the architecture and the evaluation plan for a thorough investigation of each system component.

% talk about test case - metrics
For the purposes of the evaluation, we ran a \emph{focused crawl} with a model consisting of $7$ positive and $7$ negative URLs, a sample of which may be seen on Table \ref{tab:pos_neg_examples}, and $21$ seeds extracted by the \emph{SeedFinder} component for the query ``iot vulnerabilities''. The crawl harvested around $22K$ websites per hour/per thread on a commodity machine; around $10\%$ of the frontier links were considered \emph{relevant} according to the crawler model and were thus harvested. 

%\begin{table} [t]
%	\caption{Page Relevance Metrics}
%	\centering
%	\begin{tabular}{|c|c|}
%		\hline
%		Total Pages &  257,445 \\
%		\hline
%		Relevant Pages &  26,262 \\
%		\hline
%		Irrelevant Pages &  232,183 \\
%		\hline
%		Harvest Rate &  10.201\% \\
%		\hline
%	\end{tabular}
%	\label{tab:pageRelevanceStatsTab}
%\end{table}

In order to determine the quality of information within the harvested URLs that were considered relevant in the \emph{focused crawl}, we developed a web-based \emph{evaluation tool} (a snapshot of which is shown in Figure~\ref{fig:CrawledDataEvaluation}) for collecting crowdsourced data from human experts. Using this tool, human judges are provided with a random harvested website and are asked to assess its relevance on a 4-point scale; these assessments are expected to showcase the necessity of the language models and drive our classification task (i.e., help us identify appropriate thresholding/top-k values). Early results from a limited number of judgements on the clear web crawl show that about 1\% of the harvested websites contain actionable cyber-threat intelligence, while (as expected) the percentage is higher for the social and dark web.

%The tool's purpose is to allow human judges to view a sample of the contents of each extracted website and give it a score of $1-4$, depending on whether they find it to be: . The votes can help us  This will be achieved by comparing the \emph{Relevance Score} to corresponding vote for each post in the NoSQL datastore. . According to early observations, roughly $1\%$ of the crawled data contains actionable  information. \\

Besides the evaluation tool, we have also implemented a visualisation component for the computation of the relevance score for posts. This component highlights vocabulary words on the actual crawled posts, and displays the computed relevance score according to our model. A sample output of the component for a random post is shown in Table~\ref{tab:RelevanceScoreExtraction}.

\begin{table}[!t]
%	\begin{center}
		\caption{Relevance Score Computation}
	\vspace{-3mm}
\centering
		 %excerpt from: https://www.iotforall.com/5-worst-iot-hacking-vulnerabilities/
		\label{tab:RelevanceScoreExtraction}
		\begin{tabular}{|c|c|}
			\hline
%			\multicolumn{2}{|c|}{ } \\ 
			\multicolumn{2}{|c|}{Excerpt from: \url{www.iotforall.com/5-worst-iot-hacking-vulnerabilities}} \\ 
			\hline
			\multicolumn{2}{|c|}{ } \\ 
			\multicolumn{2}{|l|}{\multirow{1}{8cm}{The \hlcyan{Mirai} \hlcyan{Botnet} (aka Dyn \hlcyan{Attack}) Back in October of 2016, the largest \hlcyan{DDoS} \hlcyan{attack} ever was launched on \hlcyan{service} provider Dyn using an \hlcyan{IoT} \hlcyan{botnet}. This lead to huge portions of the \hlcyan{internet} going down, including \hlcyan{Twitter}, the Guardian, \hlcyan{Netflix}, Reddit, and CNN.\\ 
					\bigskip
					This \hlcyan{IoT} \hlcyan{botnet} was made possible by \hlcyan{malware} called \hlcyan{Mirai}. Once \hlcyan{infected} with \hlcyan{Mirai}, computers continually search the \hlcyan{internet} for vulnerable \hlcyan{IoT} \hlcyan{devices} and then use known default \hlcyan{usernames} and \hlcyan{passwords} to \hlcyan{log} in, infecting them with \hlcyan{malware}. These \hlcyan{devices} were things like digital cameras and \hlcyan{DVR} \hlcyan{players}.}} \\
			\multicolumn{2}{|c|}{ } \\ 
			\multicolumn{2}{|c|}{ } \\
			\multicolumn{2}{|c|}{ } \\
			\multicolumn{2}{|c|}{ } \\
			\multicolumn{2}{|c|}{ } \\
			\multicolumn{2}{|c|}{ } \\
			\multicolumn{2}{|c|}{ } \\
			\multicolumn{2}{|c|}{ } \\
			\multicolumn{2}{|c|}{ } \\
			\multicolumn{2}{|c|}{ } \\ 
			\hline
			Relevance Score & 0.8563855440900794\\
			\hline
		\end{tabular}
%	\end{center}
	\vspace{-5mm}
\end{table}

%The visualization is too big to display properly and when scaled down it's unreadable. I suggest we do not show it for now.

%Figure \ref{fig:visualize2D} visualizes how the Topic and Post vectors are displayed in the 2D space after performing Dimentionality Reduction with the t-SNE algorithm.

%\begin{figure*}[!ht]
%	\centering
%	\includegraphics[scale=0.40]{figures/plot2D}
%	\caption{Classifier Output}
%	\label{fig:visualize2D}
%\end{figure*}

% !TEX root = main.tex

\section{Related Work}
\label{sec:relatedWork}

Web crawlers, typically also known as robots or spiders, are tightly connected to information gathering from online sources. %They constitute an important part of search engines and they lie at the heart of most information gathering tasks that are performed online. Since the first crawler, called the Wanderer written by M. Gray in 1993 for collecting statistics about the growth of the web, a lot of progress has been made in the field of information gathering by crawling. In this section, we outline the state-of-the-art in (Clear, Deep, and Dark) web crawling and present different aspects of the crawling technology. 
In this section we review the state-of-the-art in crawling by (i) outlining typical architectural alternatives that fit the crawling task %,(ii) categorising the available crawling solutions according to their policy for traversing the web graph, 
and (ii) categorising the different crawler types based on the type of the targeted content.

%Conceptually, the typical procedure followed by a web crawler is fundamentally simple: it visits a URL, downloads the webpage associated with it, extracts the URLs therein, compares them with a list of visited URLs and adds the non-visited ones to its frontier list (i.e., a list of URLs to be visited). This procedure is repeated until a (sub-)domain is fully crawled. Obviously, this simple procedure does not necessarily need a highly sophisticated architecture as it can nowadays be easily implemented by resorting to any of the high-level scripting languages. However, harnessing the vast scale of online resources and dealing with varying types of content and information sources requires careful engineering and involves important architectural decisions.

\subsection{Architectural typology}
\label{sec:ArchitecturalTypology}

Depending on the crawling application, the available hardware, the desired scalability properties and the ability to scale up/out the existing infrastructure, related literature provides a number of architectural alternatives \cite{Najork09, HsiehGL10, HarthUD06}.

\stitle{Centralized.} Typically, special-purpose or small-scale crawlers follow a centralised architecture \cite{Najork09}; the page downloading, the URL manipulation, and the page storage modules, resort in a single machine. This centralised architecture is, naturally, easier to implement, simpler to deploy, and straightforward to administer, but is limited to the capabilities of the hardware and thus cannot scale well. For this reason, the more sophisticated crawler designs put effort in scaling out, i.e., exploiting the inherently distributed nature of the web and adopt some form of decentralisation.
	
\stitle{Hybrid.} Hybrid crawler architectures (e.g., \cite{ShkapenyukS02}) are the norm in the architectural typology as they aim for a conceptually simple design that involves distributing some of the processes, while keeping others centralised. In such architectures, the page downloading module is typically distributed, while URL management data structures and modules are maintained at a single machine for consistency. Such designs aim at harnessing the control of a centralised architecture and the scalability of a distributed system; however the centralised component usually acts as a bottleneck for the crawling procedure and represents a single point of failure.
	
\stitle{Parallel/Distributed.} A parallel crawler \cite{Ahmadi-AbkenariS12, QuocFFRSS15} consists of multiple crawling processes (usually referred to as C-procs in crawler jargon), where each such process performs all the basic tasks of a crawler. To benefit from the parallelisation of the crawling task the frontier is typically split among the different processes, while to minimise overlap in the crawled space, links and other metadata are communicated between the processes. When all processes run on the same LAN then we refer to an intra-site parallel crawler, while when C-proc’s run at geographically distributed locations connected by a WAN (or the Internet) we refer to a distributed crawler.
	
\stitle{Peer-to-peer.} The advent of peer-to-peer computing almost two decades ago introduced peer-to-peer search engines like %YaCy\footnote{https://yacy.net/en/index.html} and 
Minerva \cite{ZimmerTW07}; this in turn gave rise to the concept of peer-to-peer crawlers \cite{VikasCRMMGS07, BambaLCPSBPPLS07}. Peer-to-peer crawlers constitute a special form of distributed crawlers that are typically targeted to be run on machines at the edge of the Internet, as opposed to their distributed counterparts that are designed for clusters and server farms. To this end, peer-to-peer crawlers are lightweight processes that emphasise crawl personalisation and demonstrate large-scale collaboration usually by means of an underlying distributed routing infrastructure \cite{VikasCRMMGS07}.%(e.g., a DHT \cite{StoicaMLKKDB03}, or a super-peer network \cite{VikasCRMMGS07}).

\stitle{Cloud-based.} Lately, the requirement for more effective use of resources by means of elasticity gave rise to a new crawler paradigm: the cloud-based crawlers \cite{GuptaMBMP16}, \cite{YanLrP14}, which revived known machinery to a renewed scope, versatility and options. Such architectures use cloud computing features alongside big data solutions like Map/Reduce and NoSQL databases, to allow for resource adaptable web crawling and serve the modern the Data-as-a-Service (DaaS) concept.

\subsection{Usage typology}
\label{sec:UsageTypology}

Although the first web crawlers that set the pathway for the spidering technology were developed for the clear (or surface) web, in the course of time specialised solutions aiming at the different facets (social, deep, dark) of the web were gradually introduced. Below we organise crawlers in terms of their intended usage.

\stitle{Clear/surface web.} Since the introduction of the first crawler in 1993, the majority of the research work on crawlers has focused on the crawling of the surface web, initially on behalf of search engines, and gradually also for other tasks. There is an abundance of work on clear web crawling; some insightful surveys include \cite{McCurley09, Najork09}.

\stitle{Web 2.0.} The advent of the user-generated content philosophy and the participatory culture of Web 2.0 sites like blogs, forums and social media, formed a new generation of specialised crawlers that focused on forum \cite{JiangYL12, YangCWHZM09, WangYLCZM08, CaiYLWZ08}, blog/microblog \cite{HurstM09, 0001S15,FerreiraLMCFL12}, and social media \cite{BuccafurriLNU12, KhanS17} spidering. The need for specialised crawlers for these websites emerged from  
(i) the content quality,
%(i) the quality (i.e., typically structured and user-generated) and creation rate of content usually found in forums/blogs, 
(ii) the inherent structure that is present in forums/blogs, %that makes it possible to even develop frameworks for creating blog crawlers 
and (iii) the implementation particularities (e.g., Javascript-generated URLs) that make other crawler types inapplicable or inefficient.

\stitle{Deep/Dark/Hidden web.} The amount of information and the inherent interest for data that reside out of reach of major search engines, hidden either behind special access websites (deep web) or anonymisation networks like Tor\footnote{\url{https://www.torproject.org/}} and I2P\footnote{\url{https://geti2p.net/}} (dark web), gave rise to specialised crawlers \cite{ValkanasNG11}. To this end, over the last ten years, a number of works related to deep web crawling have been published \cite{WangLCL17, ZhaoZNHJ16, ZhengWCJL13}, investigating also (i) different architectural \cite{ZhaoW12, LiWT12} and automation \cite{FurcheGGSS13} options, (ii) quality issues such as sampling \cite{LuWLCL08}, query-based exploration \cite{LiuWJZL09}, duplicate elimination \cite{JiangWFLZ10}, and (iii) new application domains \cite{LiWD13, HeXGRS13}.
	
\stitle{Cloud.} Finally, the elasticity of resources and the popularity of cloud-based services inspired a relatively new line of research focusing on crawler-based service configuration \cite{MenzelKLT13} and discovery in cloud environments \cite{NoorSANL13}.

\section{Outlook}
\label{sec:outlook}

We are currently working on a thorough evaluation of our architecture. Our future research plans involve the extraction of cyber-threat intelligence from the relevant harvested content, by utilising natural language understanding for named entity recognition/disambiguation.
%\input{sec6-ack}

%% Another way to reduce vertical spaces between bibliography items when using IEEEtran and compsocconf option in document class
\newcommand{\BIBdecl}{\setlength{\itemsep}{0.25 em}}

\bibliographystyle{IEEEtran}
\bibliography{crawler}

\end{document}